\documentclass[bibyear]{aa}  

\usepackage{graphicx}
\usepackage{txfonts}
\usepackage{lipsum}
\usepackage{subcaption}        
\usepackage{lscape}             
\usepackage{placeins}           
\usepackage{xcolor}
\usepackage{natbib}
\bibpunct{(}{)}{;}{a}{}{,}
\usepackage{booktabs}
\usepackage{threeparttable}
\makeatletter
\let\linenumbers\relax

\let\switchlinenumbers\relax
\let\runninglinenumbers\relax
\makeatother

\begin{document}

\title{The chemical DNA of the Magellanic Clouds}
\subtitle{IV. Unveiling extreme element production: the Eu abundance in the Small Magellanic Cloud
\thanks{Based on observations collected at the ESO-VLT under the program 113.26EF.}}

\author{
S. Anoardo\inst{1} \and
A. Mucciarelli\inst{1,2} \and
M. Palla\inst{1,2} \and
L. Santarelli\inst{1,2} \and
C. Lardo\inst{1,2} \and
D. Romano\inst{2}
}

\institute{
Dipartimento  di  Fisica  e  Astronomia  “Augusto  Righi”,  Alma  Mater  Studiorum, Universit\`a  di Bologna, Via Gobetti 93/2, I-40129 Bologna, Italy
\and
INAF - Osservatorio di Astrofisica e Scienza dello Spazio di Bologna, Via Gobetti 93/3, I-40129 Bologna, Italy
}

\date{November 12, 2025}

\abstract{
In this study we investigate the chemical enrichment of the rapid neutron-capture process 
in the Small Magellanic Cloud (SMC). We measure [Eu/Fe] abundance ratios in 209 giant stars that are confirmed members of the SMC, providing the first extensive dataset of Eu abundances in this galaxy across its full metallicity range, spanning more than 1.5 dex.
We compare Eu abundances with those of Mg and Ba to evaluate the efficiency of the r-process relative to $\alpha$-capture and s-process nucleosynthesis. 
The SMC shows enhanced [Eu/Fe] values at all metallicities (comparable with the values measured in the Milky Way), with a clear decline as [Fe/H] increases (from $\sim$ --1.75 dex to $\sim$ --0.5 dex), consistent with the onset of Type Ia supernovae. In contrast, [Eu/Mg] is enhanced by about +0.5 dex at all [Fe/H], significantly above the values observed in Milky Way stars, where [Eu/Mg] remains close to solar, reflecting comparable production of r-process and $\alpha$-capture elements. Moreover, [Ba/Eu] increases with metallicity, beginning at [Fe/H] $\approx$ --1.5 dex, namely at a lower metallicity with respect to the Milky Way, where [Ba/Eu] starts to increase around [Fe/H] $\approx$ --1 dex. 
Our findings suggest the SMC has a higher production of Eu (with respect to the $\alpha$-elements) than the Milky Way but 
in line with what observed in other dwarf systems within the Local Group.
We confirm that galaxies with star formation efficiencies lower than the Milky Way have high [Eu/$\alpha$], 
probably indicating a stronger efficiency of the delayed sources of r-process at low metallicities.

}


\maketitle

\section{Introduction}

Neutron-capture processes, overcoming the high Coulomb barriers of high-Z nuclei, 
are responsible for the formation of the majority of the elements 
heavier than the Fe-peak \citep[see e.g.][]{bbfh57}. Depending on the balance between 
the time-scales of the neutron captures and of the $\beta$-decay, 
these elements are labelled as rapid (r-), intermediate and slow (s-) neutron-capture elements. 
The r-process is the most extreme among these, needing a neutron flux of the order 
of $\sim10^{20}$ ${\rm cm^{-3}}$. 

In particular, the astrophysical sites for the occurrence of the r-process are still debated. 
Our current view is that two concurrent production sites contribute to produce 
the r-process elements: (1) a prompt source, occurring on timescale shorter than tens of Myr, involving some peculiar core-collapse 
supernovae (SNe), like the magneto-rotationally driven SNe \citep[e.g.][]{winteler12}, 
the supernova-triggering collapse of rapidly rotating massive stars
\citep[collapsars, e.g.][]{siegel19}, and highly magnetised neutron stars 
\citep[magnetars, e.g.][]{Patel25}, and (2) a delayed source, occurring 
on timescales up to some Gyr, identified in compact binary mergers (neutron star-neutron star or neutron star-black
hole, hereafter just referred to as NSMs)
corresponding to explosive kilonova events connected to the previous gravitational wave emission
\citep{argast04,thielemann17}.
The detection of electromagnetic radiation from the gravitational wave event GW170817 \citep{abbott17} 
has provided a direct evidence of the contribution by NSMs to r-process nucleosynthesis
\cite[see e.g.][]{pian17,Watson19}. 
Chemical evolution models reproduce fairly well the abundance patterns of r-process elements measured in our Galaxy by means of the combined action of both prompt and delayed sources (e.g. \citealt{Cote19,Molero23,palla2025}), further contributing to the idea of multiple sources acting on different timescales as r-process production sites.

Recently, the study of r-process elements has received considerable attention in the field of Galactic archaeology, 
in particular Europium, due to its very low "contamination" by s-process production since, in metal-poor stars, the Eu abundance pattern is consistent with a pure r-process origin \citep[see e.g.][]{Burris00}. 
In fact, the [Eu/$\alpha$] abundance ratio can serve as a powerful diagnostic for distinguishing between in-situ 
and accreted stellar populations in the Milky Way (MW) halo \citep{monty24,ernandes2024,ceccarelli2024}. 
Indeed, stars and globular clusters identified as accreted based on their dynamical properties 
exhibit higher [Eu/$\alpha$] ratios than in-situ stars at metallicities above [Fe/H]$\sim$ –1.3 dex, 
which is precisely where the largest chemical differences between in-situ and accreted populations 
are observed \citep[see e.g.][]{nissen2010,helmi2018,matsuno22,ceccarelli2024}. 
This suggests that the production of r-process elements may be higher when compared to that of $\alpha$-elements
(mostly taking place in massive stars ending their lives as core-collapse SNe, e.g. \citealt{Arcones23})
in stars formed in now-dissolved accreted systems, whose masses were comparable to those of present-day dwarf spheroidal (dSph) or ultra-faint dwarf galaxies. 

This result is consistent with studies conducted over the past two decades on both isolated dwarf galaxies and those currently interacting with or merging into the MW. In such galaxies, [$\alpha$/Fe] is typically lower than in the MW at similar [Fe/H], reflecting their lower star formation efficiency (see \citealt{Matteucci21} and references therein), while [Eu/Fe] remains elevated, indicating strong r-process activity. 
Indeed, these patterns have been observed in the Large Magellanic Cloud \citep[LMC,][]{mucciarelli10,vander13}, 
in Fornax \citep{letarte10} and Sculptor dSphs \citep{hill2019}, and in the remnant of the Sagittarius dSph \citep{sbordone2007,Reichert20,liberatori2025}. These findings underscore the importance of tracing the chemical evolution of dwarf galaxies in order to interpret the properties of accreted stellar populations now residing in the Galactic halo.

Recently, \citet{palla2025} addressed the problem to model the r-process enrichment in Local Group galaxies, by including 
both prompt and delayed sources, as commonly assumed in the literature \citep[e.g.][]{Prantzos20,Koba20,molero21}. Their results showed that chemical prescriptions able to well reproduce Eu abundance patterns in MW stars 
fail to reproduce the abundances measured in dwarf galaxies, predicting too low [Eu/Fe] and [Eu/$\alpha$] ratios, 
while are able to reproduce the measured [$\alpha$/Fe].
To solve this missing Eu problem, the authors suggested an increased production from delayed sources at low metallicity, 
which provides a much better match to the observed trends in the MW and Local Group dwarf galaxies.

All these observational and theoretical findings further underscore the need to investigate in even more detail r-process 
enrichment (and therefore Eu abundance patterns) in Local Group systems with stellar masses and star formation 
efficiencies lower than those of the MW. In this context, the Small Magellanic Cloud (SMC) 
has received very limited attention so far \citep[][hereafter Paper I]{nidever,muccia23a}, and information on the Eu abundances in its stellar populations 
remains scarce. Indeed, the only measurements of [Eu/Fe] are available for two very metal-poor SMC field stars \citep{regg21} 
and for three globular clusters at higher metallicities \citep{muccia23b}.

This work aims to fill this gap by providing, for the first time, a large dataset of Eu abundances 
derived from high-resolution spectra, in order to investigate the efficiency of r-process enrichment in the SMC relative to $\alpha$ and s-processes. The Eu abundances are also compared with those of other nucleosynthetic processes,  as the $\alpha$-capture (here traced by Mg 
and occurring in massive stars) and the s-process (traced by Ba and occurring mostly in low and intermediate-mass Asymptotic Giant Branch, hereafter AGB, stars).
In addition, Ba serves as a further indicator of the r-process efficiency at low-metallicity, as it receives significant contribution  where s-process production is disfavoured (due to the metallicity dependence of the s-process, see e.g. \citealt{Arcones23}).
In this way, 
the results presented here will provide not only new constraints on the chemical evolution of the SMC, but also offer valuable insights into the role of r-process nucleosynthesis in low-mass, low-efficiency star-forming systems within the Local Group.

The paper is organised as follows. In Section \ref{s:observations}, we summarise the observations 
and the dataset analysed in this study; Section \ref{s:spectra_analysis} describes the methods used to infer atmospheric parameters, 
radial velocities and chemical abundances of the target stars. Section~\ref{s:chem_abu} presents the results and 
Section~\ref{disc} discusses the interpretation of the measured abundance patterns. Finally, in Section \ref{s:conclusion}, we draw our conclusions.

\section{Observation and data reduction}
\label{s:observations}

This study is based on observations (Program ID:~113.26EF, PI: Mucciarelli) performed with the multi-object spectrograph FLAMES
\citep{pasquini} targeting the same fields analysed in Paper I. We focus on three SMC fields — FLD-121, FLD-339, and FLD-419 — each centred on a globular cluster (NGC 121, NGC 339, and NGC 419, respectively; see Fig. 1 in Paper I). 
These three fields are located in different positions of the SMC in order to sample the chemical composition of the SMC close to the main body of the galaxy (FLD-339 and FLD-419) and its outskirts (FLD-121). 
The observations discussed in Paper I were based on the HR11 and HR13 FLAMES setups that do not include any Eu transition in their spectral range. 

The new observations presented here were carried out with the HR15n FLAMES setup ($ \rm R=19200$, $6470$\r{A} $< \lambda < 6790$\r{A}), 
sampling the Eu II line at 6645 \r{A}. 
For each field, three exposures of 45 minutes each were secured, in order to obtain, for the co-added spectra, a signal-to-noise ratio (SNR) of $\sim$50 and $\sim$100 
for the faintest (G$\sim$16.6) and brightest (G$\sim$16) targets respectively.
A total of 209 giant stars members of the SMC were observed, 158 of them in common with the sample of Paper I 
and other new 51 targets selected from Gaia Data Release 3 photometry \citep{brown21}. 
These new stars were selected in the magnitude range G$\sim$16--16.6, excluding objects with poor photometric quality 
\citep[as traced by the Gaia parameter phot\_bp\_rp\_excess\_factor,][]{lind18}. 
This magnitude range allows to have stars with SNR larger than 50 (faint limit) and to include only stars 
belonging to the red giant branch (RGB) and older than $\sim$1Gyr (bright limit), excluding bright stars corresponding to 
younger populations. 
We also excluded stars with companions located within 2'' and brighter by $\leq$2 magnitudes in G compared to the target star, 
in order to avoid fibre contamination from nearby sources.

The spectra analysed in this work were then reduced following the ESO FLAMES-GIRAFFE pipeline\footnote{http://www.eso.org/sci/software/pipelines/}. 
All the reduced spectra were cleaned of cosmic rays, corrected for the heliocentric velocity, co-added and finally normalised. Table \ref{info} lists the main information for all the spectroscopic targets, including the Gaia identification number, the coordinates, 
the G magnitude, the identification number used in Paper I for the stars in common, the field and an indication of whether the target star is a likely binary system.

\begin{table*}[h!]
\caption{Information about the SMC spectroscopic targets.}
\label{info}
\centering
\begin{threeparttable}
\begin{tabular}{cccccccc}
\toprule
ID \textit{Gaia} DR3 & RA & Dec & G & RV & ID Paper I & Field & Binary\\
   &   (degree) & (degree) &  & (km/s) &\\
\midrule
4689845798923576704     & 6.7124313 & --71.5774614 & 16.8821 & 120.4$\pm$0.1 & FLD-121\_50 & FLD-121 & \\
4689845597059733248     & 6.7120893 & --71.5895200 & 16.7524 & 101.0$\pm$0.1 & FLD-121\_100086 & FLD-121 & \\
4689858172724094720     & 6.4880274 & --71.5481267 & 16.8272 & 136.8$\pm$0.1 & FLD-121\_100185 & FLD-121 & \\
4689852189834915072     & 6.2466252 & --71.5899642 & 16.9358 & 129.2$\pm$0.1 & FLD-121\_100211 & FLD-121 & \\
4689844978584591104     & 6.4954253 & --71.6527985 & 17.0734 & 137.5$\pm$0.1 & FLD-121\_100237 & FLD-121 & $*$\\
\bottomrule
\end{tabular}
\begin{tablenotes}
\small
\item \textbf{Notes.} Columns are: ID, coordinates and G magnitude from Gaia DR3 \citep{brown21}, measured RV from HR15n, ID as used in Paper I, field and possibility (indicated by $*$) that the target star is part of a binary system. The entire table is available at the CDS.
\end{tablenotes}
\end{threeparttable}
\end{table*}

\begin{figure}
  \resizebox{\hsize}{!}{\includegraphics{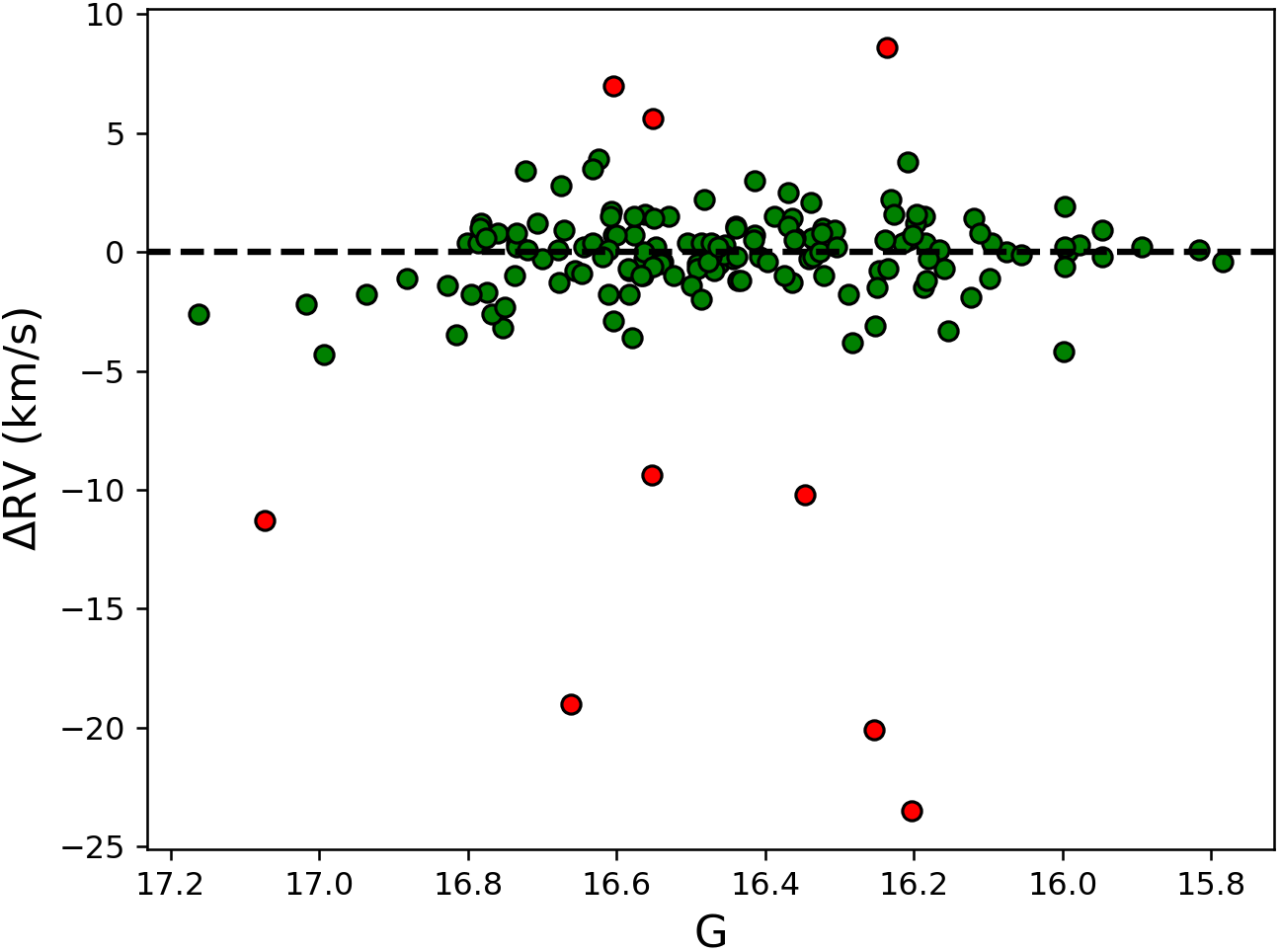}}
  \caption{Difference between the RVs derived in this study and those in Paper I as a function of 
  the Gaia DR3 G magnitude. The red circles are stars with discrepant RV differences after a 3$\sigma$-clipping iterative procedure.}
  \label{rv_SD}
\end{figure}

\section{Spectral analysis}
\label{s:spectra_analysis}

\subsection{Radial velocities}

Radial velocities (RVs) for all the new spectra were measured by exploiting the standard cross-correlation technique 
\citep[see e.g.][]{tonry} as implemented in the {\tt PyAstronomy} package.
Synthetic spectra computed with the \texttt{SYNTHE} code \citep[][]{synthe}, following the procedure described in Section 3.2, were adopted as templates.
The HR15n FLAMES setup does not include any sky emission line that allows us to check the accuracy of the wavelength calibration. 
We compared the RVs of the stars in common with Paper I (where the zero-point of the 
wavelength calibration was checked using the O emission sky line at 6300 \AA ) finding a median difference of --1 km/s. 
This offset has been applied to all the RVs derived from HR15n spectra. 
The new 51 target stars are all members of the SMC having RVs compatible with the 
RV distribution of the galaxy \citep[see e.g.][Paper I]{deleo,nidever}.
Typical uncertainties of the heliocentric RVs are $\simeq$ 0.1  km/s. 

In order to check for possible binary stars, the RVs derived in the two available epochs 
for the stars in common with Paper I have been compared (see Fig.~\ref{rv_SD}). 
Most of the data cluster around zero, with some points appearing as clear outliers, with absolute RV differences exceeding 10 km/s. 
We performed an iterative 3-$\sigma$ rejection on this sub-sample.
Convergence (i.e., no further values being excluded) is reached for a sub-sample of 149 stars, 
which can be considered as normally distributed. Nine stars are excluded as outliers and 
considered as candidate binaries. 
The surviving sample has a mean RV = -- 0.09 km/s and a standard deviation = 1.51 km/s. 
This dispersion is not compatible with the RV measurement errors, suggesting that the spread is real and 
that can be attributed to radial pulsations, 
which can cause RV variations on the order of 1–2 km/s in stars near the tip of the RGB 
\citep{carney2003,hekker2008} such as those analysed in this study.

\subsection{Atmospheric parameters}
\label{atmo}

For the stars in common with Paper I we adopted their stellar parameters, 
namely effective temperatures ($ \rm T_{eff}$), surface gravities (log~g), and microturbulent velocities ($ \rm v_t$). 
For the new targets, the stellar parameters were calculated from the photometry adopting the same approach used in Paper I. 
In particular, $ \rm T_{eff}$ were obtained from the broad-band colour $ \rm (G-K_S)_0$ using the $ \rm (G-K_S)_0 -  T_{eff}$ transformation provided by \citet{muccia_temp} and the values we calculated for $ \rm T_{eff}$ range from $\sim$3800 K to $\sim$4500 K. 
We adopted G magnitudes from Gaia Data Release 3 \citep{brown21} and $ \rm K_S$ from 2MASS \citep{skrutskie}. 
G magnitudes have been corrected for extinction following the prescriptions by \citet{gaia18}, while for $ \rm K_S$ magnitudes the extinction coefficient by \citet{mccall} was adopted. The colour excess values E(B-V) are taken from the infrared dust maps by \citet{ebv11}.

Surface gravities were obtained from the Stefan-Boltzmann law, assuming $ \rm T_{eff}$, a stellar mass equal to $ \rm 1 M_{\odot}$, a true distance modulus $ \rm DM_0=18.965 \pm 0.025$ \citep{gracz} and calculating the bolometric correction following  \citet{andrae}. 
The assumption of $ \rm 1 M_{\odot}$ is reasonable according to the age distribution of the SMC stars (see discussion in Paper I). 
While it is not possible to derive precise masses for individual targets, a variation of $ \rm \pm0.2M_{\odot}$ from this assumed value results in a change in log g smaller than 0.1 dex, with a negligible impact on the derived abundances.
When considering the distance modulus, it should be noted that the SMC is characterised by a significant line-of-sight depth that it is not easy to properly evaluate for each single target. The depth maps provided by \citet{depth} indicate that the three target fields here discussed should cover a depth range between 2 and 6 kpc. When a conservative distance variation of 3 kpc is assumed, the uncertainties in log g only increase by 0.02 dex, translating into variations of less than 0.02  in the abundances of single ionised lines (like Ba and Eu lines). The final error in log g are then dominated by the uncertainties in the stellar mass.
The values we calculated for log g range from 0.4 to 1.1 (in cgs units).

The microturbulent velocity $ \rm v_t$ is usually derived spectroscopically by removing any trend between the iron abundance and the 
strength of the lines \citep[see][for a review of this approach]{mucciarelli11}.
Our spectra have a relatively small number ($\sim$30) of FeI lines and so the spectroscopic determination of $ \rm v_t$ could be affected by statistical fluctuations leading to uncertain values of this parameter. 
We computed $ \rm v_t$ exploiting the log g - $ \rm v_t$ relation provided by \citet{muccia_boni}, which is based on the spectroscopic $ \rm v_t$ obtained from high-resolution, high-SNR spectra of giant stars in 16 Galactic GCs. The values we calculated for $ \rm v_t$ are between 1.7 and 1.9 km/s, in line with the expected values for red giant stars.

\begin{table*}[h!]
\caption{Stellar parameters and chemical abundances for the SMC spectroscopic targets.}
\label{info2}
\centering
\begin{threeparttable}
\begin{tabular}{ccccccc}
\toprule
ID \textit{Gaia} DR3 & $ \rm T_{eff}$ & log g & $ \rm v_t$ & [Fe/H] & [Eu/Fe] & [Ba/Fe] \\
   &   (K) & (cgs) & (km/s) & (dex) & (dex) & (dex)\\
\midrule
4689845798923576704     & 4319 & 1.06 & 1.7 & --1.01$\pm$0.13 & 0.23$\pm$0.08 & --0.01$\pm$0.13 \\
4689845597059733248     & 4065 & 0.84 & 1.8 & --0.89$\pm$0.11 & 0.66$\pm$0.09 & 0.27$\pm$0.13 \\
4689858172724094720     & 4084 & 0.88 & 1.8 & --1.17$\pm$0.10 & 0.86$\pm$0.11 & 0.21$\pm$0.14 \\
4689852189834915072     & 4345 & 1.09 & 1.7 & --1.14$\pm$0.14 & 0.75$\pm$0.09 & 0.03$\pm$0.13 \\
4689844978584591104     & 4293 & 1.12 & 1.7 & --0.82$\pm$0.13 & 0.42$\pm$0.09 & 0.19$\pm$0.13 \\
\bottomrule
\end{tabular}
\begin{tablenotes}
\small
\item \textbf{Notes.} Columns are: ID from Gaia DR3 \citep{brown21}, derived atmospheric parameters and abundance ratios. The entire table is available at the CDS.
\end{tablenotes}
\end{threeparttable}
\end{table*}

\subsection{Chemical analysis}
We present [Eu/Fe], [Mg/Fe] and [Ba/Fe] abundance ratios of the entire spectroscopic sample.
All the abundances were derived with our own code \texttt{SALVADOR} (D. A. Alvarez Garay et al. in prep.) that performs a $\chi^2$-minimization between the observed lines and a grid of synthetic spectra calculated by varying the abundance of the species of interest. 
The latter were calculated with the code \texttt{SYNTHE} \citep{synthe} assuming 
the appropriate stellar parameters for each target, adopting a new grid of \texttt{ATLAS9} model atmospheres 
(A. Mucciarelli et al., in prep.) and including all the atomic and molecular transitions from the compilation available in the Kurucz/Castelli linelist\footnote{https://wwwuser.oats.inaf.it/fiorella.castelli/}. All the synthetic spectra are calculated at high resolution,  
including only the intrinsic mechanisms of broadening of the lines,
and then convoluted with a Gaussian profile in order to reproduce the observed broadening.

The Eu and Ba abundances were estimated for the entire sample by measuring the Eu~II line at 6645 \AA\ (for which we adopted log gf = 0.120) 
and the Ba~II line at 6496.9 \AA\ (log gf = -- 0.407). For both lines, the synthetic spectra take 
into account the hyperfine and isotopic splittings affecting these transitions 
\citep{lawler01,NIST}. 
For the stars in common with Paper I the latter [Fe/H] values are adopted, as well as the Mg abundances, 
since there are no Mg transitions in the spectral range of HR15n. 
Also, for these stars, Ba abundances were derived in Paper I using the Ba~II line at 6141.7 \AA . 
The comparison between [Ba/Fe] values of Paper I and this study provides an average difference (this study - Paper I) of +0.22$\pm$0.02 dex. 
Therefore, we decide to rescale the values obtained from the 6496.9  \AA\ line to those of Paper I, for consistency with the previous work. 
Finally, for the 51 new targets, the Fe abundance was estimated by measuring about 30 Fe~I lines, carefully selected to be unblended given the stellar parameters, the observed spectral resolution and the chemical composition of each target.

The computation of chemical abundances is affected by two main sources of error, i.e. the uncertainties arising from the atmospheric parameters and the measurement errors due to the fitting procedure described above.
We estimated the uncertainties due to the atmospheric parameters repeating the analysis by varying each time a given parameter of the corresponding $1\sigma$ error ($ \rm \pm60 \ K$ for $ \rm T_{eff}$, $\pm0.1$ for log g and $ \rm \pm0.2 \ km/s$ for $ \rm v_t$) and keeping fixed the others.
Uncertainties related to the measurement procedure are computed as the abundance standard deviation normalised to the root mean square of the number of used lines. For Ba and Eu, for which only one transition was available, we estimated the measurement error by running Monte Carlo simulations: we created 200 synthetic spectra with a Poissonian noise that reproduces the observed SNR and then we repeated the analysis. The dispersion of the abundance distribution obtained from these synthetic spectra is assumed as the $1\sigma$ uncertainty.

The two sources of uncertainties are finally added in quadrature and since the results are expressed as abundance ratios, we took into account also the uncertainties in the Fe abundance. The final errors in the [Fe/H] and [X/Fe] abundance ratios were calculated as follows:

\begin{align}
\sigma_{\rm [Fe/H]} &= \sqrt{
\frac{\sigma_{\rm Fe}^2}{N_{\rm Fe}} + 
(\delta ^{T_{\rm eff}}_{\rm Fe})^2 + 
(\delta ^{\log g}_{\rm Fe})^2 + 
(\delta ^{v_{\rm t}}_{\rm Fe})^2
}
\end{align}

\begin{align}
\sigma_{\rm [X/Fe]} &= \sqrt{ 
\frac{\sigma_{\rm Fe}^2}{N_{\rm Fe}} + 
\frac{\sigma_{\rm X}^2}{N_{\rm X}} + 
\delta T^2 + \delta g^2 + \delta v_{t}^2
}
\end{align}

where $\delta T = \delta^{\rm T_{eff}}_X-\delta^{\rm T_{eff}}_{\rm Fe}$, $\delta g = \delta^{\rm log\ g}_{\rm X}-\delta^{\rm log \ g}_{\rm Fe}$ and $\delta v_{\rm t} = \delta^{\rm v_t}_{\rm X}-\delta^{\rm v_t}_{\rm Fe}$. $\sigma_{\rm Fe}$ and $\sigma_{\rm X}$ are the standard deviations of Fe and X respectively, $N_{\rm X}$ and $N_{\rm Fe}$ are the number of lines used to derive the abundances, and $\delta^{\rm i}_{\rm Fe}$ and $\delta^{\rm i}_{\rm X}$ are the abundance variations obtained modifying the atmospheric parameter \textit{i}.

\section{Chemical abundance ratios}
\label{s:chem_abu}

\begin{figure}
  \includegraphics[width=0.46\textwidth]{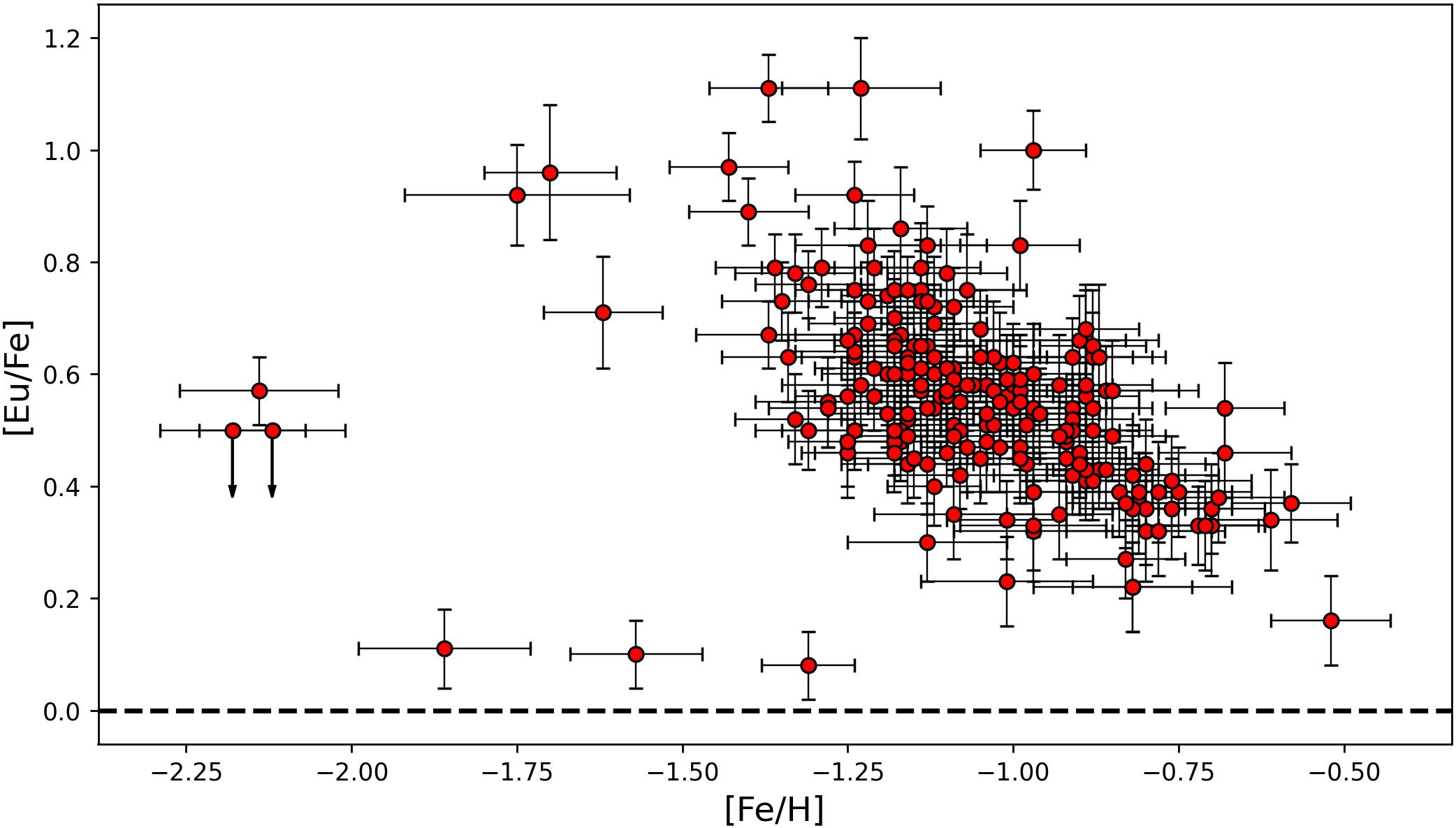}
  \includegraphics[width=0.46\textwidth]{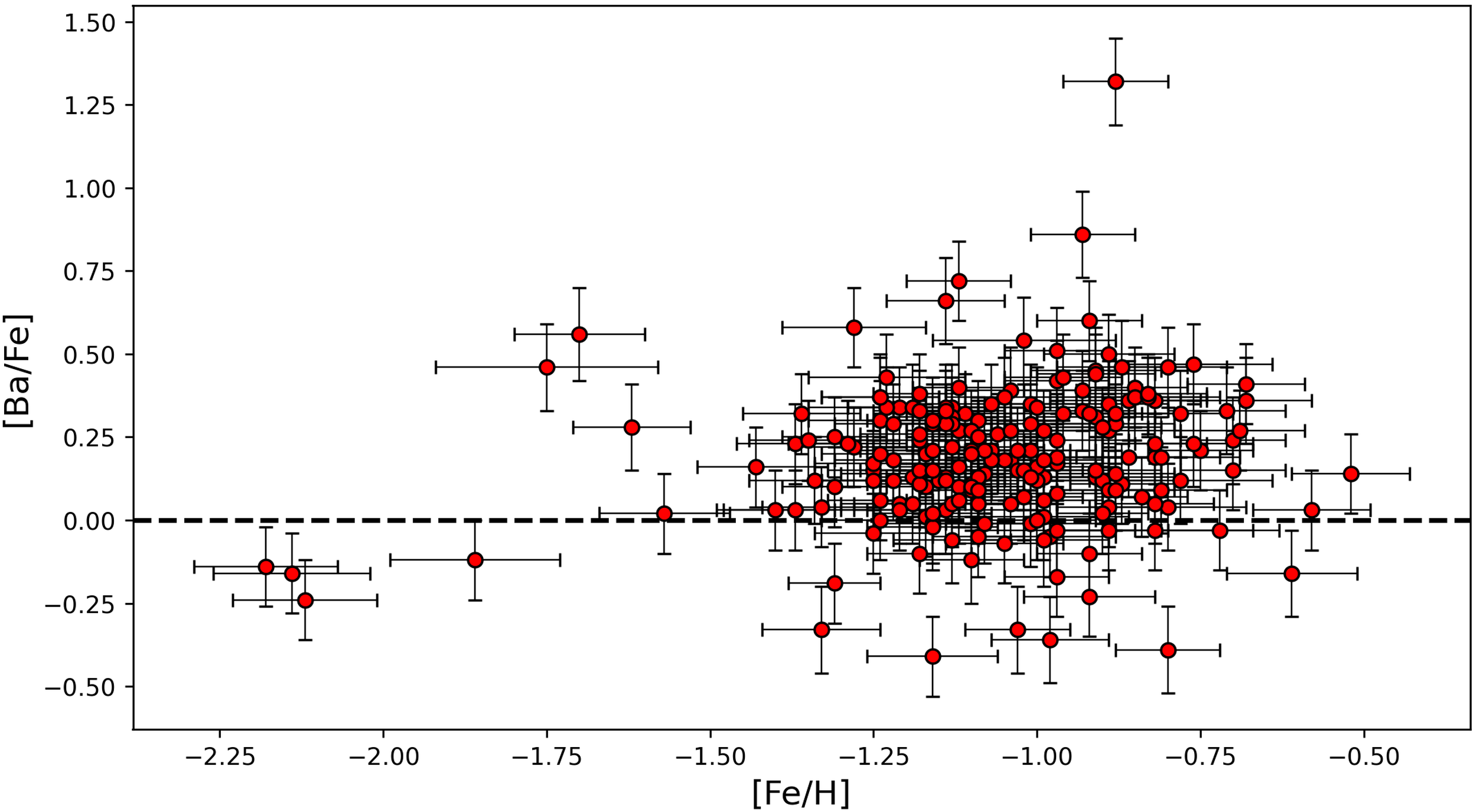}
  \caption{Abundance ratios for the analysed SMC stars: [Eu/Fe] (top panel) and [Ba/Fe] (bottom panel) as a function of [Fe/H]. 
  The horizontal dashed lines mark the solar value.}
  \label{abu_fig1}
\end{figure}

\begin{figure}
  \resizebox{\hsize}{!}{\includegraphics{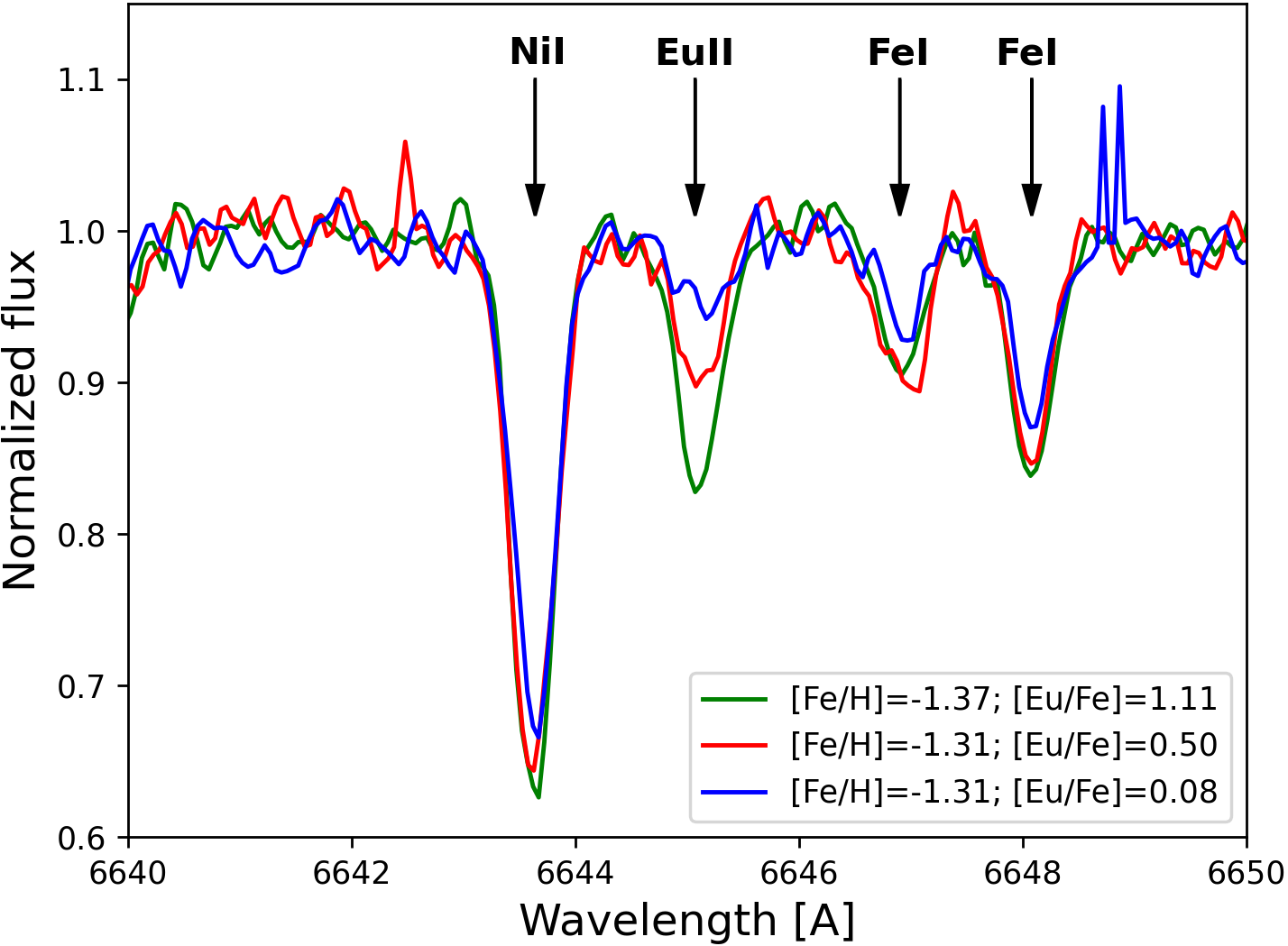}}
  \caption{Comparison between VLT-FLAMES spectra of three stars having similar metallicity and stellar parameters but different [Eu/Fe] abundances: 4685582751830393088 ($\rm T_{eff}=3881 \ K$ and log  g = 0.50, in green), 4687249169089083136 ($\rm T_{eff}=4054 \ K$ and log g = 0.45, in red) and 4684827460356570624 ($\rm T_{eff}=4087 \ K$ and log g = 0.61, in blue). The arrows mark the position of the Eu~II line at 6645 \r{A} analysed in this work and of the adjacent absorption features.}
  \label{cfr}
\end{figure}

Fig.~\ref{abu_fig1} shows the trends of [Eu/Fe] and [Ba/Fe] as a function of [Fe/H] for the 
spectroscopic sample analysed here.
All stars of our sample have supersolar [Eu/Fe] abundances, covering a range of values between $\sim$+0.2 and $\sim$+1 dex.
At metallicities below approximately --1.2 dex, the targets exhibit a pronounced star-to-star scatter in [Eu/Fe], spanning nearly 1 dex — substantially larger than the typical measurement uncertainties in the sample ($\sim$0.1 dex).
This points out the presence of an intrinsic scatter in [Eu/Fe] among the most metal-poor SMC stars. 
The existence of this scatter is readily apparent when comparing the spectra of stars with similar [Fe/H] but different [Eu/Fe]. For example, Fig. \ref{cfr} shows the spectra of three stars having similar metallicity  ([Fe/H]$\sim$--1.3 dex) and atmospheric parameters, but different strength of the Eu~II line, demonstrating an intrinsic difference in [Eu/Fe] at similar metallicity.

For metallicities above approximately $\sim$--1.2 dex, corresponding to the bulk of the sample, [Eu/Fe] exhibits instead a clear trend with [Fe/H]. 
[Eu/Fe] starts to decline for increasing [Fe/H], from [Eu/Fe]$\sim$+0.7 dex down to [Eu/Fe]$\sim$+0.35 dex. 
The analysed stars originate from three fields located at different positions within the galaxy (see Sec. \ref{s:observations}). 
As discussed in Paper I, the three fields have comparable chemical compositions but with small differences in some elements 
(i.e. Na, Ti, V and Zr) and in the fraction of metal-poor stars, suggesting a not uniform chemical enrichment history.
We verified that the [Eu/Fe] versus [Fe/H] trend is consistent across all three fields. 
The observed trends between [Eu/Fe] and [Fe/H] turn out to be very similar each other.
We carried out a Kolmogorov–Smirnov test on the possible pairs of [Eu/Fe] distributions as a function of [Fe/H]. In all cases, the resulting p-values are above 0.35, showing that the null hypothesis (that the distributions are drawn from the same parent population) cannot be rejected. Therefore, we have no hints of systematic differences in the r-process production among the three SMC fields.
Throughout the paper, we will therefore discuss the entire sample without distinguishing between the fields.  Finally, we checked the chemical abundances of the nine candidate binary stars, finding that these stars have values for all the 
abundance ratios discussed here (i.e. [Eu/Fe], [Eu/Mg], [Ba/Eu]) indistinguishable from the other stars. This points out that the possible binary nature 
of these stars does not affect their chemical composition.

Concerning [Ba/Fe], we highlight an increasing trend with [Fe/H], with the few metal-poor stars in the sample having sub-solar abundance ratios, 
and the bulk of the sample with enhanced [Ba/Fe] values, despite a large star-to-star scatter and the presence of some Ba-rich stars, similar to the findings discussed by \citet{muccia23a}. For the discussion of [Mg/Fe] as a function of [Fe/H] we refer the reader 
to \citet{muccia23a} and their Figure 8 because in this study we used their [Mg/Fe] values. Here we recall only that in that sample [Mg/Fe] is mildly enhanced ($\sim$+0.1/0.2 dex) until [Fe/H]$\sim$--1 dex, with a subsequent decrease down to subsolar [Mg/Fe] values. 
In general, [Mg/Fe] in the SMC stars results to be lower than MW stars of similar [Fe/H].

\begin{figure}
  \includegraphics[scale=0.38]{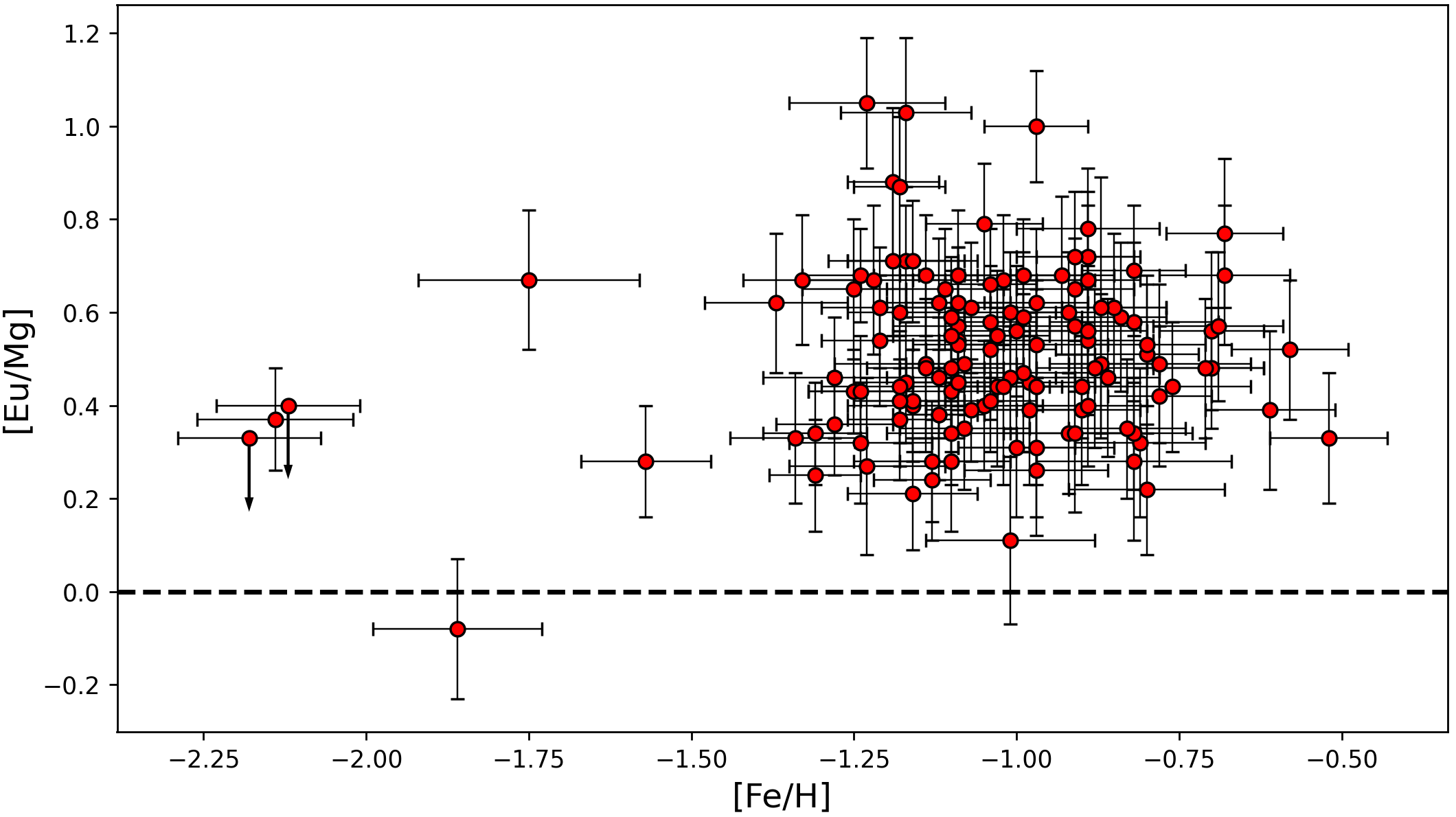}
  \caption{[Eu/Mg] as a function of [Fe/H] for the spectroscopic sample discussed in this study.}
  \label{abu_fig2}
\end{figure}

In Fig. \ref{abu_fig2} we show the observed trend for SMC stars in [Eu/Mg]. This abundance provides an excellent indicator of the relative efficiency of the r-process production compared to the $\alpha$-capture in massive stars. Notably, Mg is produced exclusively in massive stars, unlike other $\alpha$-elements (e.g. Si, S, Ca) which receive non-negligible contribution by Type Ia SNe (e.g. \citealt{Palla21}).
The SMC stars consistently show (except one star, Gaia DR3 4689844153950876928) super-solar values in [Eu/Mg]. Overall, the trend remains approximately flat across the metallicity range, albeit with a non-negligible scatter ($\sigma$= 0.17 dex), and is centred around [Eu/Mg] $\sim$ +0.5 dex.

\begin{figure}
  \includegraphics[scale=0.38]{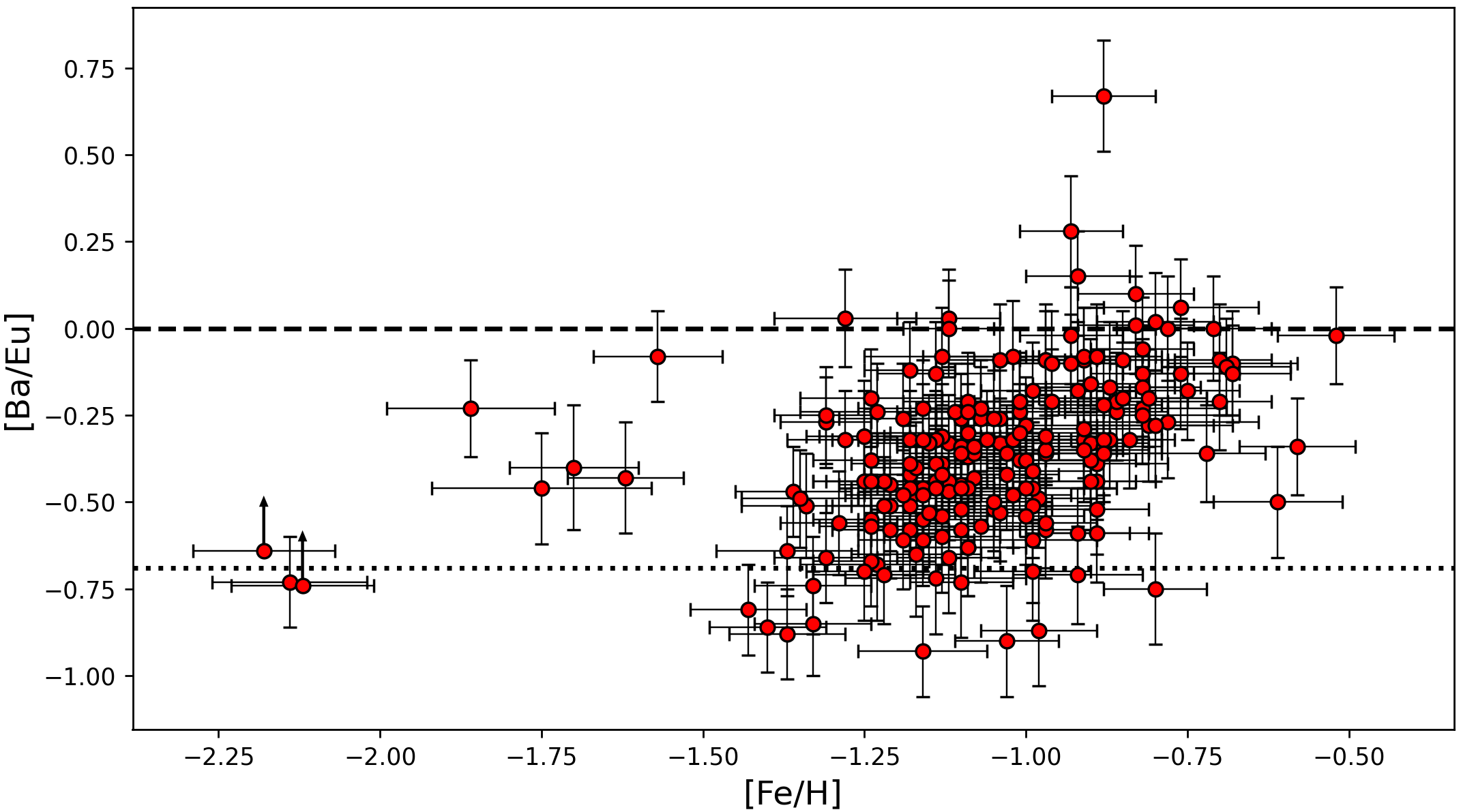}
  \caption{[Ba/Eu] as a function of [Fe/H] for the spectroscopic sample discussed in this study. 
  The black dotted line in the bottom panel marks [Ba/Eu]=--0.69 dex, indicative of pure r-process content in the stars \citep{arla}.}
  \label{abu_fig3}
\end{figure}

Finally, in Fig. \ref{abu_fig3}, we display the [Ba/Eu] versus [Fe/H] abundance trend for our sample of SMC stars. 
Such a diagram reveals the relative contributions to Ba production from the s- and r-processes as a function of metallicity. Ba is an element mostly produced by the s-process (85\%  at solar metallicity, e.g. \citealt{Sneden08}), but a small amount is also produced by the r-process, with a ratio relative to Eu of [Ba/Eu]=--0.69 dex for the  solar r-process pattern (\citealt{arla}): the dotted horizontal line in the bottom panel of Fig. \ref{abu_fig3} indicates this solar pattern.
Our abundance ratios show a  scattered behaviour from the few stars in the low-metal end of the sample (up to [Fe/H]$\sim$--1.5 dex), with values ranging from  [Ba/Eu]$\sim$--0.5 dex to solar.
For the bulk of the stars at higher metallicity, instead, we see a well-defined increase in [Ba/Eu] with [Fe/H]. In particular, we observe a rise from [Ba/Eu]$\sim$--0.6 dex, namely close to the expected outcome from r-process production only, up to [Ba/Eu]$\sim$0.2 dex, indicating a progressive increase of the s-process contribution with metallicity.

\begin{figure}
  \resizebox{\hsize}{!}{\includegraphics{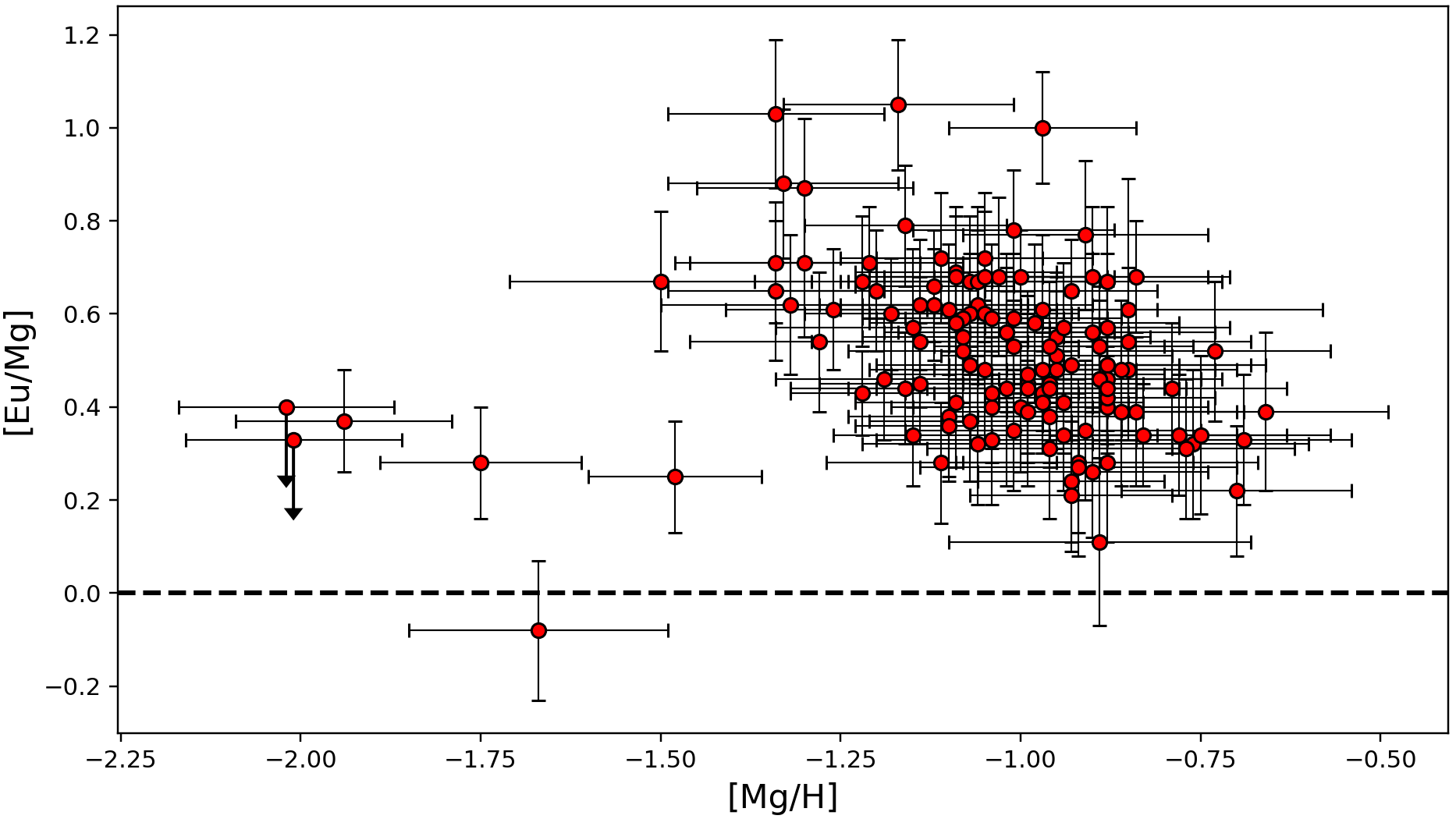}}
  \resizebox{\hsize}{!}{\includegraphics{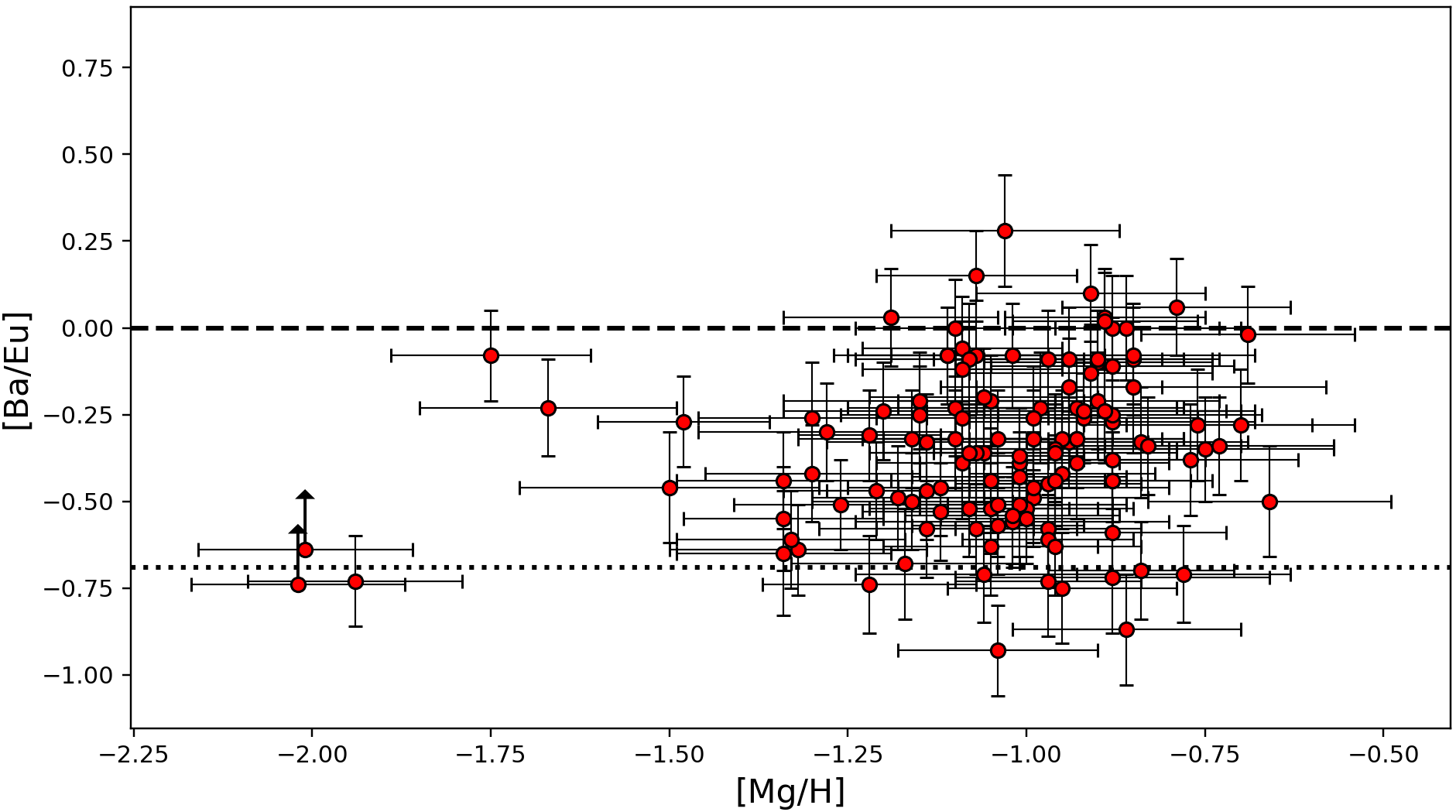}}
  \caption{[Eu/Mg] (top panel) and [Ba/Eu] (bottom panel) as a function of [Mg/H].}
  \label{abu_fig4}
\end{figure}

Additionally, we investigated the behaviour of [Eu/Mg] and [Ba/Eu], two abundance ratios useful to quantify the 
efficiency of r-processes relatively to $\alpha$- and s-processes, as a function of [Mg/H] (see Fig.~\ref{abu_fig4}). 
In fact, [Mg/H] is often used as metallicity scale alternative to [Fe/H]
as Mg is produced almost exclusively by short-lived massive stars (core-collapse SNe), while Fe has significant delayed contributions from Type Ia SNe; 
thus [Mg/H] traces early, SN II-dominated enrichment more cleanly than [Fe/H] 
\citep[among the studies proposing this approach, see e.g.][]{shig98,cayrel04}.
Interestingly, [Eu/Mg] shows a decreasing trend with [Mg/H] that mirrors that observed when we use Fe as reference, 
while the same abundance ratio appears to be almost constant when plotted as a function of [Fe/H] 
(see Fig.~\ref{abu_fig3}). On the other hand, [Ba/Eu] shows an increasing trend with [Mg/H] similar 
(but less evident) to that observed with [Fe/H]. It is worth noting that the these trends 
are less defined than those of Fig.~\ref{abu_fig2} and ~\ref{abu_fig3} because [Mg/H] is measured 
by one only transition and therefore less accurately defined than [Fe/H].

\section{Discussion}
\label{disc}

The observed [Eu/Fe] versus [Fe/H] abundance pattern in SMC as shown in Fig. \ref{abu_fig1} upper panel can be viewed in the light of the standard framework used to interpret the Eu abundance trends in the MW.
At low metallicities, despite the low number of targets prevents us to draw definitive conclusions, the large star-to-star scatter seems to suggest an inhomogeneous mixing in the gas of the SMC in its early epochs. 
A similar [Eu/Fe] scatter is observed also among the MW Halo stars \citep[see e.g.][]{mcw1995,ryan96,Burris00,francois2007,francois24}.
As discussed in Paper I, the SMC stars with [Fe/H]$<$--1.3 dex were likely formed in the first 1-2 Gyr of life of the galaxy 
\citep[see, e.g. the theoretical age-metallicity relation by][]{Pagel98}. As the production of Eu is dominated by rare sources, with both prompt and delayed sources of enrichment constituting a small minority in massive stars ($m \gtrsim 8 M_\odot$, already small in number due to the power-law nature of the initial stellar mass function, IMF), the observed star-to-star scatter reflects the stochastic nature of these sites of production (e.g. \citealt{Cescutti15,Cavallo21,Hirai22}). 
At higher metallicities, a clear decreasing trend in [Eu/Fe] with [Fe/H] is instead found.  
This can be 
attributed to the onset of the Type Ia SNe that produce Fe but not Eu, thus reducing [Eu/Fe]. 
Indeed, this progressive decrease in [Eu/Fe] is in theoretical agreement with the flat trend in [Eu/Mg] (Fig. \ref{abu_fig2}), as Mg is not produced by Type Ia SNe but only in massive stars \citep[see e.g.][]{Palla21}. 
On the other hand, the decline of [Eu/Mg] as a function of [Mg/H] suggests that increasing the bulk of the metals (traced by oxygen, which shares common timescale of enrichment to Mg) the 
relative contribution of r- and $\alpha$-processes decreases, therefore indicating of the reduction of the Eu production in massive stellar populations. Nonetheless, as already stated in Section \ref{s:chem_abu}, some caution is needed in evaluating this trend, as due to the prominent uncertainties for the Mg abundances.
At the same time, the [Ba/Eu] trend increases with metallicity, passing from low values, compatible with the theoretical expectations for pure r-process production \citep{arla,Sneden08}, to supersolar values. 
The increase of [Ba/Eu] with metallicity, starting at [Fe/H] $\sim$ -- 1 dex in the solar vicinity and at lower [Fe/H] values in the Magellanic Clouds, is readily explained as due to the delayed release of s-process elements in the interstellar medium by AGB stars (e.g. \citealt{Karakas10,Cristallo15}), occurring at lower metallicities in smaller galaxies (see below). 
Indeed, as [Fe/H] increases, the contribution from low- and intermediate-mass AGB stars eventually becomes the main production channel for Ba.

To better understand the chemical abundance trends displayed in Fig.~\ref{abu_fig1},~\ref{abu_fig2} and ~\ref{abu_fig3}, it is crucial to compare the abundance patterns of [Eu/Fe], [Eu/Mg] and [Ba/Eu] as a function of [Fe/H] as measured in the SMC stars with those observed in other systems. 
In this context, the MW and the LMC are the two most natural candidates to look at. The MW serves as the standard benchmark for chemical abundance trends
within the Local Group, while the LMC, as a companion galaxy to the SMC, occupies an intermediate scale in galactic mass between the SMC and the MW \citep{stanimirovic2004,erkal2019,callingham2019}.

\begin{figure}
  \resizebox{\hsize}{!}{\includegraphics{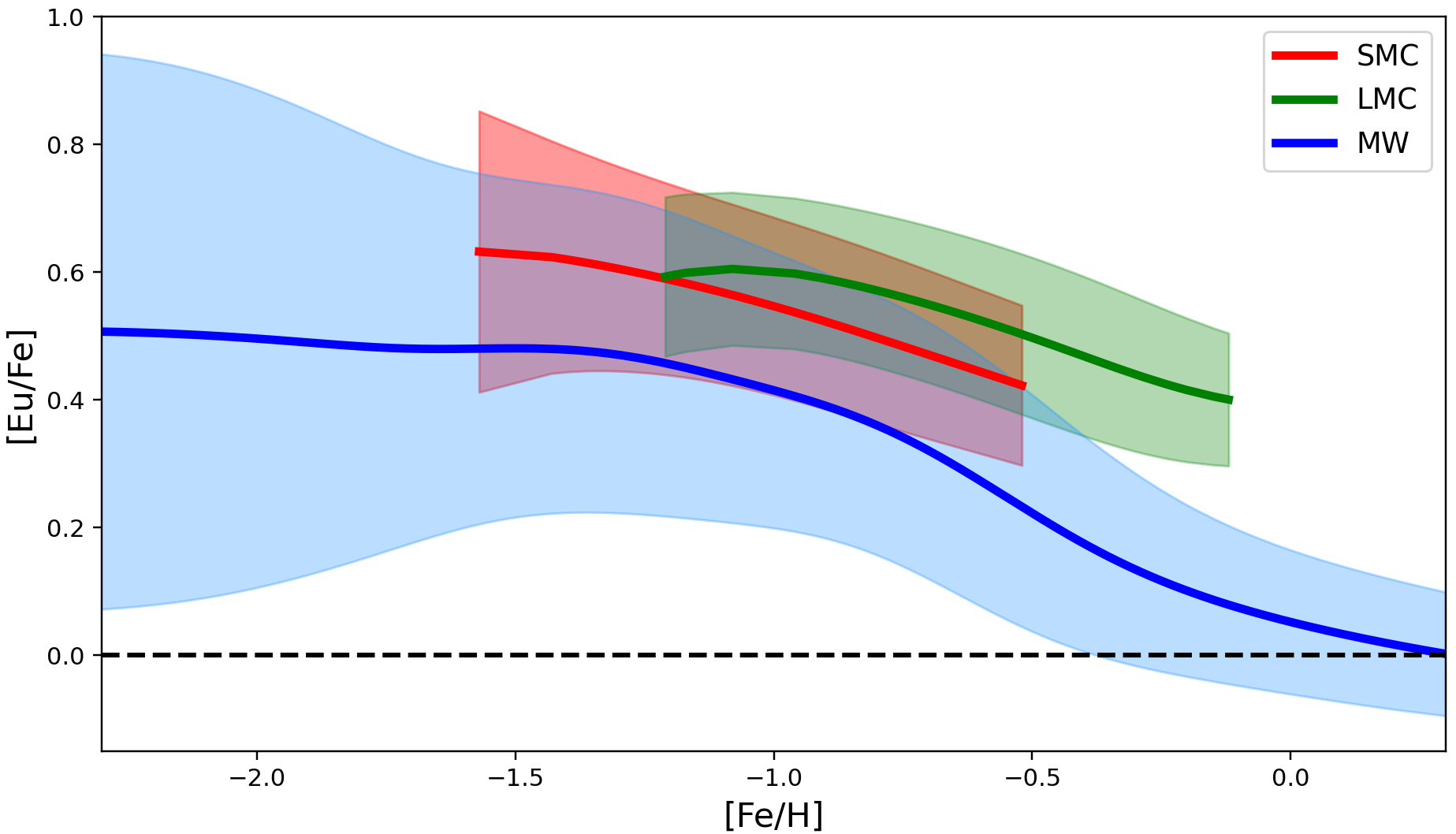}}
  \resizebox{\hsize}{!}{\includegraphics{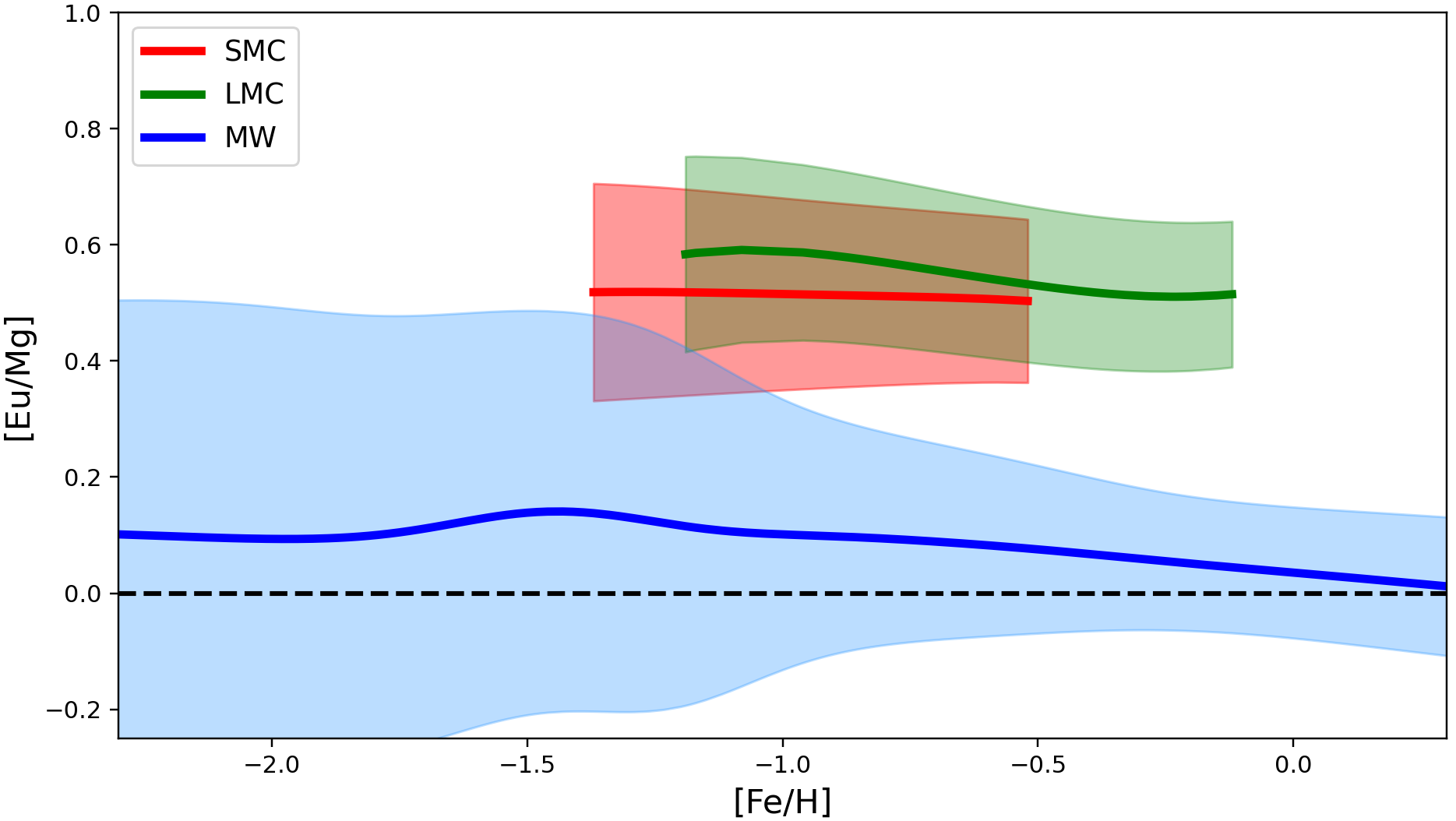}}
  \resizebox{\hsize}{!}{\includegraphics{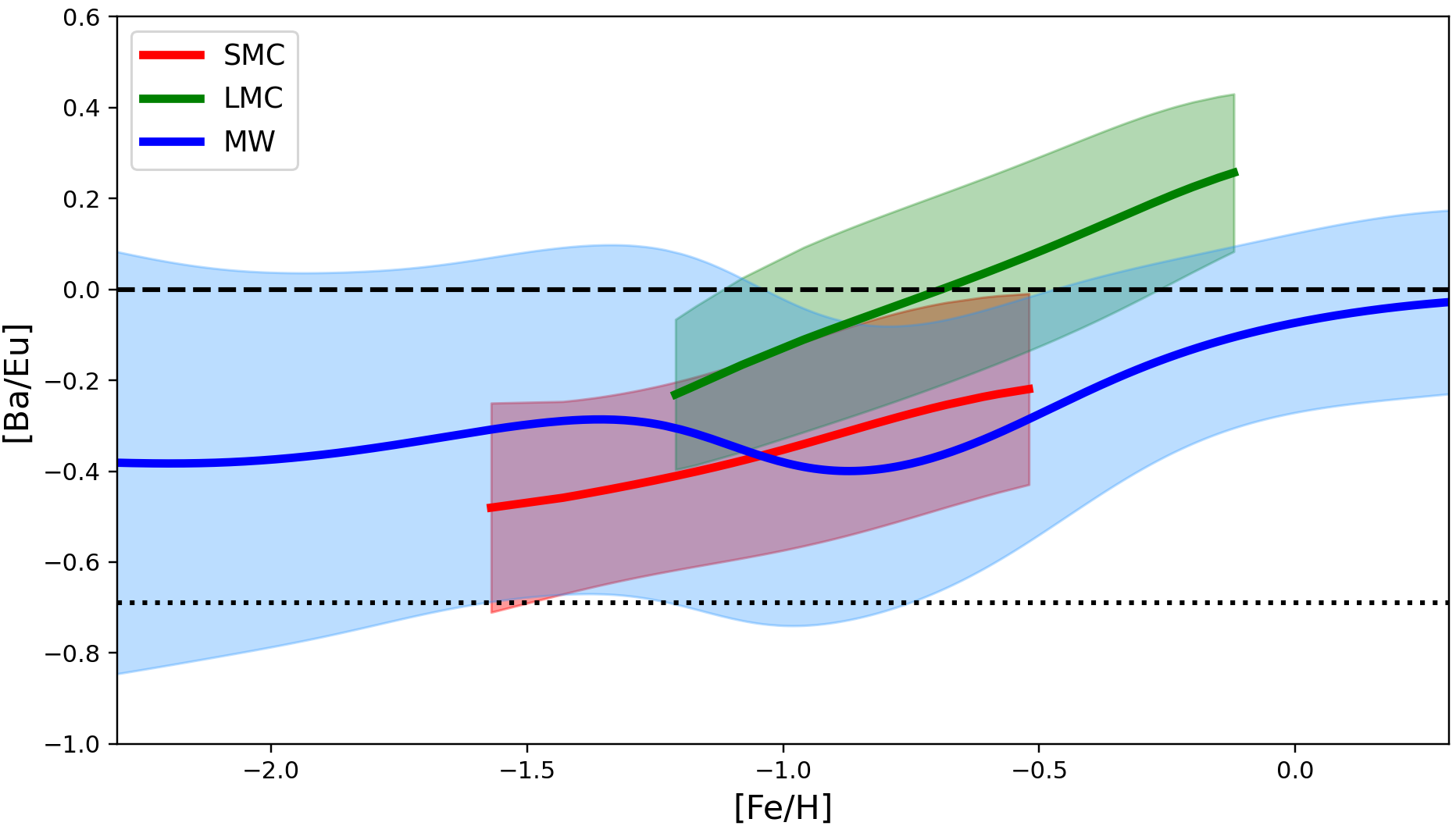}}
  \caption{[Eu/Fe] (top panel), [Eu/Mg] (middle panel) and [Ba/Eu] (bottom panel) vs. [Fe/H] trends for the SMC, the LMC and the MW. The solid lines are non-parametric Gaussian KDE regressions, while the shaded areas represent the 1$\sigma$ confidence interval of the regressions. The black dotted line in the bottom panel marks the value below which we have a pure r-process enrichment \citep[][see also Fig. \ref{abu_fig3}]{arla}.}
  \label{trend}
\end{figure}

Therefore, in Fig. \ref{trend} we show the abundance trends in our SMC stars as compared with the observations in the MW and in the LMC.
To compare our observed trends with those of the MW and the LMC, we consider the compilation of abundances for MW stars available in the SAGA database \citep{suda08} and the sample of LMC giant stars analysed by \citet{vander13}. 
The mean trends of the three galaxies (solid lines in Fig. \ref{trend}) are represented using a non-parametric Gaussian KDE regression, together with their corresponding 1$\sigma$ confidence intervals (shaded areas).
Regarding [Eu/Fe] (Fig. \ref{trend}, top panel), the SMC and LMC exhibit similar trends, with comparable median [Eu/Fe] in the metallicity range in common between the two galaxies (--1.2$\lesssim$[Fe/H]/dex$\lesssim$--0.6 ), suggesting similar timescales for the r-process in the two Clouds.
Both galaxies, however, show [Eu/Fe] values on average higher than 0.2 dex with respect to the MW stars at similar [Fe/H].  
The most pronounced difference between the trends observed in the Magellanic Clouds and that of the Galaxy is seen for [Eu/Mg], displayed in the central panel of Fig. \ref{trend}. Indeed, the SMC and LMC show comparable values in abundance ratios as happens for [Eu/Fe], but here their [Eu/Mg] reveal a much more significant offset (by $\sim$0.5 dex) than the trend observed for MW stars. 
Similar to [Eu/Fe], the comparable trends of [Eu/Mg] in the SMC and the LMC point out similar timescales 
also for $\alpha$-processes in the two Clouds.\\
For the [Ba/Eu] abundance pattern (Fig. \ref{trend}, bottom panel), we also see that the abundance ratio is generally enhanced in the Magellanic Clouds relative to the MW at metallicities [Fe/H]$\gtrsim-1$ dex, with the [Ba/Eu] in the high-metallicity end of the LMC distribution larger by $\sim$0.3 dex than that observed in the MW. However, in this panel we note that rather than an offset at all the overlapping metallicities (as in the case of [Eu/Mg]), the difference in the [Ba/Eu] abundance is the result of a progressive detachment of the SMC and LMC abundance trends
from that of the MW with increasing metallicity.

The analysis of Fig. \ref{trend} clearly highlights an excess in r-process (and therefore of Eu) production in the SMC relative to the Galaxy. 
Indeed, both panels showing [Eu/Fe] and [Eu/Mg] display values in the SMC comparable with those in the LMC, but enhanced with respect to what is observed in the MW, especially for [Eu/Mg].
This result is consistent with what has been observed in other Local Group galaxies with a star formation efficiency less pronounced than that of the MW, for instance the remnant of the Sagittarius dSph \citep{liberatori2025},  Fornax \citep{letarte10} and Sculptor dSphs \citep{hill2019}. 
All these galaxies appear to share high [Eu/Fe] and simultaneously low [$\alpha$/Fe] values, resulting in [Eu/$\alpha$] abundance ratios being markedly different between the MW and smaller galaxies (see also \citealt{palla2025}). 

Moreover, this is also consistent with the evidence that stars and globular clusters in the MW dynamically associated with merger events involving primitive, satellite galaxies are characterised, at least for [Fe/H]$>$--1.5 dex, by higher [Eu/$\alpha$] abundance ratios compared to in-situ stars 
\citep{monty24,ernandes2024,ceccarelli2024}. 
Therefore, the findings in this work point towards an ubiquitous r-process enhancement across MW satellite galaxies, from the most massive systems (LMC and SMC) to least massive dSphs (Sculptor), urging further theoretical work to explain with a self-consistent theory the r-process abundance patterns observed throughout the Local Group. On this regard, \citet{palla2025} suggested an increased production in r-process from NSM/delayed sources for Z$\lesssim 0.1$ Z$_\odot$ to match the enhanced [Eu/Fe] and [Eu/$\alpha$] observed in several Local Group dSphs (Sagittarius, Fornax, Sculptor) with the observed trends in the MW. 
A similar scenario, but based on an increased production in r-process from prompted sources, has been proposed by \citet{tsujimoto24} 
to explain the observed Eu abundances in the LMC.
It is worth noting that the observed pattern in the [Eu/Mg] vs. [Mg/H] diagram is line with this claim, as Mg traces the enrichment timescale of the bulk of metals, and a progressive decrease in the r-process to $\alpha$-element ratio should mirror a fading contribution by an r-process source.

However, further investigations are definitely needed in this context, also considering that theoretical predictions have to match observations not only for Eu and metals up to the iron-peak  (e.g. Fe, Mg) but also for other neutron-capture elements, which receive a non-negligible contribution from the r-process to their production.
On this regard, Fig. \ref{trend} bottom panel is highly instructive, showing the differences in the trends of [Ba/Eu] in the SMC, LMC and the Galaxy at different metallicities. 
At variance with what happens for, e.g. [Eu/Mg], here we note a rather "standard" behaviour for the [Ba/Eu] abundance ratio, with an enhancement in Ba at lower metallicties for systems with lower stellar mass and star formation efficiency. Indeed, the contribution expected from the first low-mass AGB stars (producing s-process) is expected to take place at lower metallicities in the latter systems \citep{venn04,tolstoy2009} where the chemical enrichment proceed at slower pace (\citealt{matte2012}).

The stronger production of r-process elements up to [Fe/H]$\sim$--1 dex proposed by \citet{palla2025}, however, is going in the opposite direction, as it suggests to decrease the relative s-process contribution to Ba production, therefore leading to a flatter trend.
A possible solution can be seen in the context of a different IMF in dwarf galaxies, with the IMF of the Magellanic clouds being more bottom-heavy / top-light, therefore balancing the overproduction of Eu with more Ba production by low-mass AGB stars. This solution well agrees to the multiple indications \citep[e.g.][]{Lee09,Pflamm09,Jerabk18,Watts18} pointing towards a reduction in the relative fraction of massive stars in dwarf galaxies stellar populations. 
However, a thorough discussion on this matter would go beyond the scope of this paper, and will be addressed in future paper of this series
(Palla et al. in prep.).

\section{Summary and conclusions}
\label{s:conclusion}

In this study we present the first measurements of the [Eu/Fe] ratio in a sample of SMC stars covering a very extended (1.5 dex in [Fe/H]) 
metallicity range of the galaxy. 
The most interesting feature that we found from this dataset is the enhanced [Eu/Mg] in the SMC, 
comparable with the values measured in the LMC and in other dSphs but significantly higher than the MW values. 
Our result confirms that the Local Group galaxies characterised by a star formation efficiency lower than than that of the MW 
(i.e. Sagittarius, isolated dSphs, Magellanic Clouds) exhibit [Eu/Fe] and [Eu/Mg] clearly distinct with respect to the MW stars, 
therefore supporting the idea that [Eu/$\alpha$] can be used as a powerful diagnostic to distinguish accreted stars from those formed in situ \citep{monty24,ernandes2024,ceccarelli2024}, at least at intermediate-/high-metallicity. 
On the other hand, the SMC exhibits a standard behaviour for [Ba/Eu] that increases by increasing [Fe/H], 
indicating a progressive increase of the contribution of low- and intermediate-mass AGB stars with the age.
The investigation of neutron-capture elements in external galaxies, like the Magellanic Clouds, 
is therefore a field of research fundamental for the Galactic archaeology and for our understanding of the 
MW assembly history.

\section*{Data availability}
Tables 1 and 2 are only available in electronic form at the CDS via anonymous ftp to cdsarc.u-strasbg.fr (130.79.128.5) or via http://cdsweb.u-strasbg.fr/cgi-bin/qcat?J/A+A/.

\begin{acknowledgements}

We thank the referee, Sten Hasselquist, for the useful comments and suggestions. A.M., M.P. and D.R. acknowledge support from the project "LEGO – Reconstructing the building blocks of the Galaxy by chemical tagging" (P.I. A. Mucciarelli) granted by the Italian MUR through contract PRIN 2022LLP8TK\_001. 

\end{acknowledgements}

\bibliographystyle{aa}
\bibliography{biblio}

\end{document}